\documentclass[12pt,preprint]{aastex}
\begin {document}
\title{{\bf Simulated Versus Observed Cluster Eccentricity Evolution}}
\author{Stephen N. Floor\altaffilmark{1,3}, Adrian
L. Melott\altaffilmark{1} and Patrick M. Motl\altaffilmark{2}}
\altaffiltext{1}{Department of Physics \& Astronomy, University of
Kansas, Lawrence, KS 66045} \altaffiltext{2}{Department of Physics \&
Astronomy, Center for Astrophysics and Space Astronomy, Boulder, CO
80309} \altaffiltext{3}{snfloor@ku.edu}

\newpage
\begin{abstract}
The rate of galaxy cluster eccentricity evolution is useful in
understanding large scale structure.  Rapid evolution for $z < $ 0.13
has been found in two different observed cluster samples.  We present
an analysis of projections of 41 clusters produced in hydrodynamic
simulations augmented with radiative cooling and 43 clusters from
adiabatic simulations.  This new, larger set of simulated clusters
strengthens the claims of previous eccentricity studies.  We find very
slow evolution in simulated clusters, significantly different from the
reported rates of observational eccentricity evolution.  We estimate
the rate of change of eccentricity with redshift and compare the rates
between simulated and observed clusters.  We also use a variable
aperture radius to compute the eccentricity, r$_{200}$.  This method
is much more robust than the fixed aperture radius used in previous
studies.  Apparently radiative cooling does not change cluster
morphology on scales large enough to alter eccentricity.  The
discrepancy between simulated and observed cluster eccentricity
remains.  Observational bias or incomplete physics in simulations must
be present to produce halos that evolve so differently.
\end{abstract}
\newpage

\section{Introduction}

One would expect eccentricity evolution of an isolated galaxy cluster
due to violent relaxation of the system (Aarseth \& Binney 1978).  It
has been proposed (Melott et al. 2001, Plionis 2002) that one can put
constraints on $\Omega_m$ by measuring the rate of morphological
changes in clusters.  Eccentricity can measure these changes since it
is usually measured on the outer regions of clusters.

Melott, Chambers, and Miller (2001, hereafter MCM) reviewed five
observational cluster data sets (three optical and two X-ray) and
found evolution in each case with varying significance.  Plionis
(2002, hereafter PL02) also found significant evolution in cluster
eccentricity in the optical APM cluster catalog.  Recently, Floor et
al. (2003, hereafter FMMB) presented findings that showed much slower
evolution of eccentricity in simulated clusters.  It is possible that
the addition of radiative cooling might produce simulated clusters
that better emulate observed clusters.  The hydrodynamic simulations
presented in FMMB have now been outfitted to include radiative cooling
(see Motl et al. 2003).  We also have a larger sample of simulated
clusters which improves our statistics greatly.

Galaxy clusters are potentially useful for studying the nonlinear
growth of density perturbations.  Eccentricity measurements of
clusters aid in understanding the growth of clusters and large scale
structure.  Our procedure emphasizes the outer regions of clusters and
is not particularly sensitive to small scale changes in cluster core
density.  For this reason, eccentricity evolution provides a means to
measure changes in cluster morphology on the largest scales.  Also,
eccentricity presents a valuable tool to observational cosmology since
loss of data on small scales will not drastically affect the measured
value.

The question posed is whether radiative cooling will reduce the
disagreement between simulated and observed cluster eccentricity.
Cooling of the central gas of a cluster would cause contraction
followed by deepening of the central potential well.  This could cause
dark matter to preferentially reside in the center yielding a lower
eccentricity over time.  Therefore, one might expect the introduction
of radiative cooling to increase the rate of simulated eccentricity
evolution.  However, if the change in morphology due to cooling is on
small scales then this might not affect the eccentricity of the outer
regions, since the cooling time at the outer regions of clusters is
long.  The cooling time in the core of clusters is small which causes
a collapse of the region.  Since the outer regions are hydrostatically
supported by the inner regions one would expect that cooling in the
center could cause baryonic infall.  This moving gas will perturb the
dark matter potential which could cause extra-core morphology
changes. Regardless of the effects of cooling on individual clusters,
when mergers take place the accreted substructure can change
significantly when cooling is applied (Motl et al. 2003).  For these
and pedagogical reasons an investigation of the effects of radiative
cooling on galaxy cluster eccentricity evolution is presented.

The paper proceeds as follows: in \S2 we discuss the simulations used
to produce our result.  The method of eccentricity computation is
discussed in \S3.  The results and discussion are in \S4 and \S5,
respectively.  Acknowledgments are listed in \S6.

\section{Simulations}

There were two sets of simulations used in FMMB.  One is an N-Body
code with only dark matter which of course has no radiative cooling.
The hydrodynamic simulations being analyzed here were conducted with a
coupled Adaptive Mesh Refinement Eulerian hydrodynamics \& N-body code
(Norman \& Bryan 1999, Bryan, Abel \& Norman 2000).  The baryonic
fluid is evolved with the Piecewise Parabolic Method (Colella \&
Woodward 1984) and the dark matter particle potential is calculated
with an adaptive particle mesh scheme using the second-order accurate
TSC interpolation.  Each individual cluster simulation evolves the
same cosmological volume (with box length 256 Mpc) with periodic
boundary conditions and deploys the adaptive mesh infrastructure about
a different region of interest.  Each cluster region is statically
refined by two nested grids and within the innermost static subgrid,
further subgrids are created as needed to track collapsing regions.
The dark matter particles exist on the three static grids and have a
peak mass resolution of $1.3 \times 10^{10} M_{\odot}$.  For the
calculations presented here, each subgrid is refined by a factor of
two compared to its parent and we allow up to seven levels of
refinement yielding a peak spatial resolution of 16 kpc.

We use a flat $\Lambda$CDM cosmology with the following parameters:
$\Omega_{b} = 0.026$, $\Omega_{m} = 0.3$ and $\Omega_{\Lambda} = 0.7$
and we assume a Hubble constant H$_{0} = h$ 100 km s$^{-1}$ Mpc$^{-1}$
with $h =$ 0.7.  Our initial conditions are generated with the
Eisenstein \& Hu (1999) form for the CDM power spectrum and we use a
normalization of $\sigma_{8} = 0.93$.  The two samples of numerical
clusters presented here derive from the same initial conditions and
cosmological model and differ only in that for one sample the baryonic
fluid is allowed to lose energy to radiation and cool.  The adiabatic
sample will be known as $\Lambda$CDMH and the cooled sample as
$\Lambda$CDMRC.  We use a tabulated cooling curve for a plasma of
fixed metal abundance of 0.3 solar and the cooling curve is truncated
at a minimum temperature of $10^{4} K$ (Westbury \& Henriksen 1992).
For the present work we neglect the effects of thermal conduction as
well as star formation and supernova feedback.

We have investigated the morphological effects of radiative cooling on
the $\Lambda$CDMH simulations.  For a complete description of
radiative cooling see Motl et al. (2003).  We used clusters that were
isolated using the ``HOP'' algorithm (Eisenstein \& Hut 1998).  This
algorithm is based on overdensities and will select all regions above
some threshold density and then merges them based on other
considerations.  The full procedure is discussed in FMMB and
Eisenstein \& Hut (1998) but note that we used the following parameter
set: $\delta_{peak}$ = 480, $\delta_{saddle}$ = 400, and
$\delta_{outer}$ = 160.  Clusters were detected in three-dimensional
space while analysis will be done in two-dimensional projection.  We
chose to analyze only the most massive clusters in this simulation.
Due to the large volume of the simulated region (256$^3$ Mpc$^3$) all
clusters detected in this fashion were richness $R \geq 2$.  However,
FMMB used clusters richness $R \geq 1$ so conclusions from these data
seem to be richness independent.  

Simultaneously, a larger set of the $\Lambda$CDMH simulations than
available to FMMB were prepared and made available for analysis.  In
FMMB the results of eleven analyzed clusters were presented.  Here,
thirty more projected clusters were simulated using the same
parameters.  This gives us much better statistics than in FMMB.

\section{Measuring Eccentricity}

Due to widely varying definitions of ellipticity ($\epsilon$) we chose
instead to use the mathematically defined term eccentricity.  For an
ellipse with major axis $a$ and minor axis $b$, the eccentricity ($e$)
is defined to be
\begin{eqnarray}
e = \sqrt[]{1 - \frac{b^2}{a^2}}
\end{eqnarray}
All results are presented in terms of $e$.  Note that some authors use
this formula to define ellipticity, resulting in some of the ambiguity
of the term.

Since we are using three-dimensional simulations it would be possible
to measure three-dimensional features of the isolated clusters.
However, to better emulate observed results we chose to analyze the
clusters in projection.  Therefore, three projections of each cluster
were made, one along each Cartesian axis.  Jing \& Suto (2002) discuss
a robust triaxial halo morphology measurement technique.  This method
works well but the halos are identified in three dimensions.  This was
not possible in the observational data used here; moreover a method of
applying the triaxial halo analysis to observational data is not
presented in Jing \& Suto (2002).  Lee \& Suto (2004) present a
deprojection technique based on Sunyaev-Zel'dovich or X-ray
observational data.  This method is however reliant on a
five-parameter fit and commonly results in errors of 20 percent or
higher.  It is doubtful that any benefit in accuracy in eccentricity
computation could be garnished given that the inertia tensor method is
not wildly inaccurate.  A full study should be conducted
regarding the difference in eccentricity of projected clusters using
the inertia tensor method presented here with deprojected 2D clusters
using the method of Lee \& Suto (2004).  However, since the
observational data used in MCM and PL02 come from both optical and X-ray sources
it is impossible to deproject them all and the triaxial halo method is therefore
discounted.

To compare with previous
observational studies we analyze both projected dark matter and
simulated projected X-ray emissions.  We create synthetic X-ray images by projecting the
calculated X-ray emission (both line and free-free emission) from the
gas assuming a metal abundance of 0.3 relative to solar and in an
energy band extending from 0.5 to 2.0 keV.  We assume that
the dark matter halos are representative of the observed
optical emissions.  
This can be justified by noting that optical emissions come
primarily from galaxies which are approximately collisionless bodies.
The relaxation of galaxies and dark matter is therefore assumed to be
similar.  White et al. (1993) discuss the distribution of baryonic
matter in galaxy clusters.  They determine that the baryon fraction
only deviates significantly from the cosmological baryon fraction near
the core of clusters.  This implies that for the remainder of the
cluster, the region where our aggregate measures are being conducted,
the baryonic fraction is comparable to the cosmological value.
Additionally, Mellier (1999) reviews many studies where weak lensing
is used to show that the dark matter distribution in galaxy clusters is similar to both
the optical and X-ray emission distributions.  This is only true if the $n^2$
dependence of X-ray emissions are taken into account.

We identify the cluster center as the center of mass of the objects
produced by the group finding algorithm.  A common substitute for the
center of mass is to choose the highest luminosity pixel of an image.
Here this would correspond to the highest density point in the
cluster, assuming that luminosity $\propto$ density.  This procedure
was implemented to see if it affects the measured eccentricities.  The
change in eccentricity was rarely larger than the standard deviation
of the sample.  We continue to use the center of mass because it is
more easily specified and more robust to variation in observational
details.

A brief study was conducted to determine the typical displacement of
the highest density peak is from the center of mass.  The highest
density peak was determined by first rebinning the projected clusters
into larger bins to remove statistical fluctuations.  The bins were
selected such that one side is approximately 100 kpc (at $z$ = 0.1),
or comparable resolution to X-ray surveys.  The center of the bin with
the highest density was noted.  It is assumed that this point would be
analogous to the highest luminosity point of an observed cluster.  The
mean separation between the center of mass and density peak was 0.32
Mpc with $\sigma_{sep} =$ 0.37 Mpc.  The separation of the two is
quite erratic; often if the cluster has significant substructure the
highest density point will occur in an outlier of the cluster while
the center of mass is closer to the center of the projected region.
This study was conducted using the dark matter density as a measure.
Observational studies obviously cannot use the dark matter density so
another brief study was conducted.  The brightest X-ray pixel was
discovered using the same methodology as above.  The X-ray luminosity
appears to locate the dark matter center of mass somewhat better than
either the highest density dark matter or baryonic pixel.  When the
dark matter center of mass was compared to the brightest X-ray pixel
the mean separation was 0.30 with $\sigma_{sep} =$ 0.25 Mpc.  We
therefore advocate using the brightest X-ray pixel as the center of an
image rather than the brightest optical pixel, when both are
available.

Once the cluster center is determined, we emulated procedures used in
the studies discussed in MCM and PL02.  A circle of radius r$_{outer}$
= 1.5 h$^{-1}$ Mpc is drawn about the cluster center.  This circle is
commonly referred to as the aperture radius.  The mass of the material
inside the aperture radius is then determined.  An inner circle is
then drawn such that twenty percent of the mass is contained in the
annulus.  It was the eccentricity of this annulus that was measured.
See FMMB for a brief discussion of this method as compared to others.
We also used the virial radius, r$_{200}$ as the aperture radius.
This radius is defined to be the radius of a circle in which the
density is 200 times the background density of the simulation.  In
this case it was always computed in two dimensions, though it could in
principle be done in three.  Results using this as the aperture radius
are presented along with the fixed radius.

Once the annulus is isolated, the moment of inertia about the cluster
center is computed.  It is only the material that is inside the
annulus that is used to compute the moment of inertia.  The
eigenvalues of the moment of inertia tensor are proportional to the
square of the object's axes.  We therefore measure the eccentricity as
follows:
\begin{eqnarray}
e = \sqrt[]{1 - \frac{\lambda_1}{\lambda_2}}
\end{eqnarray}
with $\lambda_1 < \lambda_2$.  This method's correctness is subject to
its application to a homogeneous ellipse.  Obviously, clusters are not
entirely homogeneous objects but this is the best known method of
determining the eccentricity.

\section{Results}

In Table 1 the results for the $\Lambda$CDMH simulations are
presented.  Table 2 is analogous except for the slightly smaller size
data set of the $\Lambda$CDMRC simulations.  For each cluster we show
the median, mean, and standard deviation in the mean ($\sigma_e$) of
the eccentricity.  Inspection of tables 1-4 and their corresponding
$\sigma_e$ values will reveal that they are nearly identical.  Tables
3 and 4 display the same information as 1 and 2 except using r$_{200}$
as r$_{outer}$.  Table 5 presents calculated slopes (\emph{de/dz}) for
both the simulational and observational data sets.  The slopes and
errors ($\sigma_s$) were calculated using a least squares algorithm.
The observational data sets are described in either MCM or PL02 as
indicated.  While all slopes indicated are larger than zero, the
observational slopes are always much larger.  Before the observational
slopes were calculated the ellipticities ($\epsilon$) were converted
to eccentricities.  The standard definition of $\epsilon$
\begin{equation}
\epsilon = 1 - \frac{b}{a}
\end{equation}
with a and b as defined previously.  Both MCM and PL02 reported
results in this form.

Note that in every case presented the data has some positive slope.
The observational data has a slope close to one while simulational
slope is always much less than one.  This supports the conclusion seen
in FMMB: simulated clusters evolve slower than observed clusters.
This result was checked with the recalculation of eccentricity subject
to r$_{outer}$ = r$_{200}$.  Despite a reduction in eccentricity for
all simulated clusters, the rate of evolution remains unchanged.  It
seems that the r$_{outer}$ choice was not the source of the
discrepancy.

{\footnotesize
\begin{deluxetable}{lccccc}
\tablewidth{0pt} \tablecaption{$\Lambda$CDMH Cluster Eccentricities,
$N$=129, $R > 2$} \tablehead{ \colhead{} & \colhead{} & \colhead{} &
\colhead{} & \colhead{} & \colhead{} \\ \cline{1-6} \\
\colhead{Simulation type} & \colhead{$\sigma_8$} & \colhead{Redshift}
&\colhead{ Median $e$} & \colhead{Mean $e$} &
\colhead{$\sigma_e$(mean)} } \startdata $\Lambda$CDMH (mass) & 0.93 &
0.  &.65 & .63 &.012 \\ $\Lambda$CDMH (mass) & -- & 0.1 &.67 & .67
&.012 \\ $\Lambda$CDMH (mass) & -- & 0.25 &.72 & .70 &.012 \\
$\Lambda$CDMH (X-ray) & 0.93 & 0.  &.65 & .66 &.015 \\ $\Lambda$CDMH
(X-ray) & -- & 0.1 &.71 & .70 &.016 \\ $\Lambda$CDMH (X-ray) & -- &
0.25 &.73 & .72 &.015 \\ \enddata
\end{deluxetable}}

{\footnotesize
\begin{deluxetable}{lccccc}
\tablewidth{0pt} \tablecaption{$\Lambda$CDMRC Cluster Eccentricities,
$N$=123, $R > 2$} \tablehead{ \colhead{} & \colhead{} & \colhead{} &
\colhead{} & \colhead{} & \colhead{} \\ \cline{1-6} \\
\colhead{Simulation type} & \colhead{$\sigma_8$} & \colhead{Redshift}
&\colhead{ Median $e$} & \colhead{Mean $e$} &
\colhead{$\sigma_e$(mean)} } \startdata $\Lambda$CDMRC (mass) & 0.93 &
0.  &.67 & .64 &.012 \\ $\Lambda$CDMRC (mass) & -- & 0.1 &.69 & .68
&.012 \\ $\Lambda$CDMRC (mass) & -- & 0.25 &.70 & .69 &.012 \\
$\Lambda$CDMRC (X-ray) & 0.93 & 0.  &.67 & .68 &.016 \\ $\Lambda$CDMRC
(X-ray) & -- & 0.1 &.73 & .71 &.015 \\ $\Lambda$CDMRC (X-ray) & -- &
0.25 &.74 & .73 &.015 \\ \enddata
\end{deluxetable}}

{\footnotesize
\begin{deluxetable}{lccccc}
\tablewidth{0pt} \tablecaption{$\Lambda$CDMH Cluster Eccentricities
(r$_{200}$), $N$=129, $R > 2$} \tablehead{ \colhead{} & \colhead{} &
\colhead{} & \colhead{} & \colhead{} & \colhead{} \\ \cline{1-6} \\
\colhead{Simulation type} & \colhead{$\sigma_8$} & \colhead{Redshift}
&\colhead{ Median $e$} & \colhead{Mean $e$} &
\colhead{$\sigma_e$(mean)} } \startdata $\Lambda$CDMH (mass) & 0.93 &
0.  &.61 & .60 &.012 \\ $\Lambda$CDMH (mass) & -- & 0.1 &.64 & .64
&.011 \\ $\Lambda$CDMH (mass) & -- & 0.25 &.68 & .66 &.012 \\
$\Lambda$CDMH (X-ray) & 0.93 & 0.  &.65 & .64 &.015 \\ $\Lambda$CDMH
(X-ray) & -- & 0.1 &.64 & .63 &.015 \\ $\Lambda$CDMH (X-ray) & -- &
0.25 &.69 & .68 &.015 \\ \enddata
\end{deluxetable}}

{\footnotesize
\begin{deluxetable}{lccccc}
\tablewidth{0pt} \tablecaption{$\Lambda$CDMRC Cluster Eccentricities
(r$_{200}$), $N$=123, $R > 2$} \tablehead{ \colhead{} & \colhead{} &
\colhead{} & \colhead{} & \colhead{} & \colhead{} \\ \cline{1-6} \\
\colhead{Simulation type} & \colhead{$\sigma_8$} & \colhead{Redshift}
&\colhead{ Median $e$} & \colhead{Mean $e$} &
\colhead{$\sigma_e$(mean)} } \startdata $\Lambda$CDMRC (mass) & 0.93 &
0.  &.63 & .61 &.012 \\ $\Lambda$CDMRC (mass) & -- & 0.1 &.65 & .63
&.012 \\ $\Lambda$CDMRC (mass) & -- & 0.25 &.68 & .66 &.012 \\
$\Lambda$CDMRC (X-ray) & 0.93 & 0.  &.67 & .67 &.016 \\ $\Lambda$CDMRC
(X-ray) & -- & 0.1 &.70 & .70 &.015 \\ $\Lambda$CDMRC (X-ray) & -- &
0.25 &.67 & .67 &.015 \\ \enddata
\end{deluxetable}}

{\footnotesize
\begin{deluxetable}{rccc}
\tablewidth{0pt} \tablecaption{Summary of Eccentricity Evolution for
Observational and Simulated Data} \tablehead{ \colhead{} & \colhead{}
& \colhead{} & \colhead{} \\ \cline{1-4} \\ \colhead{Data Set (Paper
Source)} & \colhead{\large$\frac{de}{dz}$} & \colhead{$\sigma_s$}
&\colhead{ $N$}} \startdata Optical (MCM \& PL02) & 1.07 & 0.14 & 497
\\ X-ray (MCM) & 1.13 & 0.48 & 48 \\ Adiabatic Hydrodynamic Sim DM &
0.27 & 0.07 & 387 \\ Adiabatic Hydrodynamic Sim X-ray & 0.24 & 0.09 &
387 \\ Cooled Hydrodynamic Sim DM & 0.19 & 0.07 & 369 \\ Cooled
Hydrodynamic Sim X-ray & 0.18 & 0.09 & 369 \\ CDM N-Body Sim (FMMB low
$\Omega$) & 0.56 & 0.13 & 162 \\

\enddata
\end{deluxetable}}

\section{Discussion}

As discussed in FMMB we checked our conclusions by varying many
parameters related to these simulations.  We varied the value of
$\sigma_8$, the random seed for density perturbations, the cluster
detection algorithm and its parameters, the presence and magnitude of
a cosmological constant, and others.  We have now also explored the
influence of radiative cooling, changing the aperture radius, and the
definition of the center of a cluster.  
None of these alterations drastically changed the results from
previous eccentricity studies.  Changing the aperture radius reduced the
inter-simulation discrepancy.  However, the rate of evolution remains
significantly slower in simulated clusters than in observed ones, even
with the introduction of radiative cooling in the simulations.  We
discuss here the inter-simulation disagreement, the disagreement
between observation and simulation, possible sources of error, and
future work that may aid in understanding this problem.

Using r$_{200}$ as the aperture radius significantly reduced the
inter-simulation eccentricity difference.  We propose that this value
be used for the aperture radius in future studies since it is not
difficult to compute and produces more physically correct results.
When working with simulations it is quite easy to compute numerically
in projection or otherwise.  Alternately, observational studies can
use the analytic result presented in Navarro, Frenk, and White (1997).
While r$_{200}$ is not precise it does reflect the variation between
clusters of different mass and allows for a more consistent
measurement of eccentricity.

There are two main differences noticed between simulated and observed
clusters.  The first is the actual measured eccentricity.  In
simulated clusters the average eccentricity is about 0.6 or higher
which is 0.37 in ellipticity.  Observed clusters at $z \approx$ 0.05
have ellipticities of around 0.3, suggesting that present day clusters
would have lower eccentricity than the predicted present day simulated
clusters.  Therefore, lack of evolution notwithstanding, there is a
problem with simulated cluster morphology.  The greater discrepancy is
the rate of eccentricity evolution as presented in FMMB.  Radiative
cooling did not significantly change the rate of change of
eccentricity.  The change seen in \emph{de/dz} in Table 5 suggests
even slower evolution in a radiatively cooled simulation.  Changing
the aperture radius did not significantly change the eccentricity
evolution of any sample.  Either simulations
are lacking critical physics which cause much faster cluster
relaxation or observational cluster samples are missing either high
eccentricity clusters at low $z$ or low eccentricity clusters at high
$z$.

We briefly discuss two physical processes currently being integrated
into cosmological simulations that may help to alleviate this
discrepancy, simulated star formation and thermal conduction.
Simulated star formation (and subsequent supernovae) would tend to
heat the gas inside the cluster.  However, it is quite difficult to
implement a simulation which has detail down to single-star levels and
can simultaneously simulate a cosmological volume.  Further, any
morphological effects from star formation would be present at high $z$
and result only in the production of a lower density core.  It is
doubtful that star formation would change the shape of clusters on
large enough scales to alter the measured eccentricity significantly.
See Bryan et al. (2001) for a complete description of a star formation
model.  Additionally, as suggested by Narayan and Medvedev (2001) and
Medvedev et al. (2003), thermal conduction from the hot outside of the
cluster to the center of the cluster can be critical to the X-ray
emissions of a cluster.  Loeb (2002) discusses that while conductivity
may not be responsible for cooling core clusters, it can affect the
temperature distribution in clusters.  He shows that a large
heat conduction coefficient leads to cooling of the cluster gas which
transitively affects the intergalactic medium.  These temperature
changes outside the
core could easily affect the X-ray morphology of clusters.  Any flows
via conductivity can affect the eccentricity of clusters over time since
these flows involve the transfer of energy across a significant
distance.  Thermal conductivity should be added to simulations to
better emulate reality but it is again doubtful that it will change
the eccentricity significantly.  There is no obvious missing physics
which should drastically alter the morphology of simulated clusters.

Observational bias or incompleteness in the currently available
cluster catalogs could also produce the observed discrepancy.  MCM
studies the ellipticity evolution in observational samples.  The
various samples were cut to only include those clusters which were
members of the MX galaxy survey.  Miller et al. (1999) have shown that
this sample has few projection effects and that only $\sim 5\%$ of
clusters are spurious detections of overdensities on the sky.
Additionally, the co-moving number density of these clusters is nearly
constant to $z = 0.1$, indicating its level of completeness for
systems $R \ge 1$.  However, selection effects present in the various
studies, both optical and X-ray, may have persisted in spite of this
cut.  Highly eccentric, low $z$ clusters could be hard to identify
using standard cluster detection algorithms due to their large spread
on the sky.  We turn to two new observational surveys which will
hopefully add to the completeness of observational cluster catalogs.

We anticipate results from the SDSS (Nichol et al. 2000) which should
be completed shortly.  This catalog promises to be quite complete and,
provided a proper algorithm is chosen for cluster detection,
bias-free.  We feel that the C4 algorithm discussed in Nichol et
al. (2003) appears to be a non-biased cluster detection algorithm
applied to a complete sample of galaxies.  Assuming that optical
emissions are a tracer of dark matter, a comparison between simulated
dark matter density and optical emissions will possibly shed light on
the presented discrepancy.  Additionally, we look forward to a new
cluster catalog derived from the XMM-Newton survey.  XMM-Newton has
good spatial resolution and excellent sensitivity (Arnaud et al. 2002)
making it a good candidate for dim cluster measurements.  Inclusion of
these dim clusters could help the discrepancy presented.  Simulated
X-ray emission is well understood and is not subject to the assumption
that it directly traces dark matter as optical emissions are.  
Other new X-ray surveys will also make this result more robust.

\section{Acknowledgments}

We thank C. Miller for helpful discussion, and M. Plionis and
S. W. Chambers who provided data from previous papers.  SNF and ALM 
gratefully acknowledge the support of the 
National Science Foundation through grant AST-0070702, especially a
supplement for Research Experiences for Undergraduates.  Invaluable
computing support came from the National Center for Supercomputing
Applications.

\vfill \eject

\end {document}